**A theoretical-numerical model for the peeling of elastic membranes**

Daniele Liprandi[1], Federico Bosia[2], Nicola M. Pugno[1,3,4*]

[1] Laboratory of Bio-Inspired and Graphene Nanomechanics, Department of Civil, Environmental and Mechanical Engineering, University of Trento, Via Mesiano 77, I-38123 Trento, Italy

[2] Department of Physics and Nanostructured Interfaces and Surfaces Interdepartmental Centre, Università di Torino, Via P. Giuria 1, 10125 Torino, Italy

[3] School of Engineering and Materials Science, Queen Mary University, Mile End Rd, London E1 4NS, UK

[4] KET Labs, Edoardo Amaldi Foundation, Via del Politecnico snc, 00133 Rome, Italy

*nicola.pugno@unitn.it

**Abstract**

The adhesive behaviour of biological attachment structures such as spider web anchorages is usually studied using single or multiple peeling models involving "tapes", i.e. one-dimensional contacts elements. This is an oversimplification for many practical problems, since the actual delamination process requires the modelling of complex two-dimensional adhesive elements. To achieve this, we develop a theoretical-numerical approach to simulate the detachment of an elastic membrane of finite size from a substrate, using a 3D cohesive law. The model is validated using existing analytical results for simple geometries, and then applied in a series of parametric studies. Results show how the pull-off force can be tuned or optimized by varying different geometrical or mechanical parameters in various loading scenarios, and the length of the detachment boundary, known as the peeling line, emerges as the key factor to maximize adhesion. The approach presented here can allow a better understanding of the mechanical behaviour of biological adhesives with complex geometries or with material anisotropies, highlighting the interaction between the stress distributions at the interface and in the membrane itself.

## 1. Introduction

Adhesion is a topic that has attracted great interest in the mechanics community in recent years. The field of biological materials has allowed to exploit theories for adhesion formulated in the past years (Kendall, 1975; Maugis, 1992; Palacio and Bhushan, 2012) and has stimulated the formulation of novel theories and models for complex problems emerging from bio-mimetics (Lai et al., 2009; Carbone et al., 2011; Prokopovich and Starov, 2011; Brodoceanu et al., 2016; Cutkosky, 2015), from bio-mechanics (Arzt et al., 2003; Tian et al., 2006; Grawe et al., 2014; Labonte and Federle, 2016) or even from nano-mechanics (Rakshit and Sivasankar, 2014; Mo et al., 2015). Biological adhesives have been studied in depth for the optimization process they have undergone in the course of thousands of years of evolution (B. Chen et al., 2009; Pugno and Lepore, 2008; Wolff and Gorb, 2016). The term "smart adhesion" has been introduced to describe the amazing adhesive properties common to different species of animals and plants (Bhushan, 2007; Brely et al., 2018a), which have been a source of inspiration for structures for adhesive elements and manipulators in robotics (Kim et al., 2008; Daltorio et al., 2005). Frictional properties of adhesive systems have also been recently discussed (Shen et al., 2009; Das et al., 2015; Tian et al., 2006), and considerable steps have been made in tribology to investigate the behaviour observed at the small scale, leading to new adhesion, adhesion-friction and adhesion-wear models (Leonard et al., 2012; Menga et al., 2018; Vakis et al., 2018). This is often achieved by modelling the interface between the body and the substrate using elements governed by a traction-displacement law (Dimaki et al., 2016). This feature is the basis of Cohesive Zone Models (CZM) (Barenblatt, 1962; Xu and Needleman, 1994; Dimitri et al., 2015; Park and Paulino, 2013), which have been recently used to analyse the interaction between adhesion and friction (Salehani et al., 2018).

In the literature, adhesive problems are mainly described by referring to two configurations: contact mode and peeling mode, which are based on the Johnson-Kendall-Roberts (Johnson et al., 1971) and Derjaguin-Muller-Toporov (Derjaguin et al., 1975) theories, and on the Kendall single peeling theory, respectively. The considered geometries exploit symmetries to derive 1D or 2D solutions. Recent works have shown how the Boundary Element Method can be used to numerically solve adhesive problems for an arbitrarily-shaped contact area between an elastic half-space and a rigid indenter (Pohrt and Popov, 2015; Rey et al., 2017). However, these models only treat normal contact problems, where the indenter is applied vertically, and thus the adhesive directionality of the membrane is not analysed. In general, the problem of describing how an elastic membrane of finite size adheres, deforms and delaminates from an adhesive surface remains to be fully addressed. The solution of this problem is of interest both for fundamental mechanics and biology, as well as for applications in areas like the biomedical or packaging sectors.

In this paper, we propose a three-dimensional approach which combines a lattice model (Ostoja-Starzewski, 2002; H. Chen et al., 2014; Brely et al., 2016) and a CZM to describe the adhesive properties of elastic membranes. Solutions are sought for varying geometries, loading conditions and membrane properties, including anisotropy, so as to include as subcases known results in the literature, such as tape single peeling and axisymmetric membrane peeling.

## 2. Model

### 2.1 Interface

Delamination processes are often simulated using CZM. These are based on traction-separation laws, i.e. cohesive laws, which simulate the behaviour of an adhesive interface (Dugdale, 1960; Barenblatt, 1962; Park and Paulino, 2013). It was shown (Savkoor and Briggs, 1977; Warrior et al., 2003; McGarry et al., 2014) that in adhesive contact problems detachment occurs in a mixed-mode configuration and a coupled cohesive law is necessary, in which the traction along the *i*-th direction for every single node of the membrane depends upon its displacement along all 3 direction components. Despite the extensive literature on the subject, most cohesive laws are two-dimensional and only a few works deal with 3D cohesive zones. A widespread practice is to avoid a complete definition of a 3D cohesive law by using an effective gap value

$$\Delta_{eff} = \sqrt{\Delta_z^2 + \beta^2\Delta_x^2 + \beta^2\Delta_y^2} \tag{1}$$

where the fracture propagation line is assumed to belong to the $xy$ plane and $\beta$ is a scalar value used to assign different weights to the normal gap $\Delta_z$ and the tangential gaps $\Delta_x$ and $\Delta_y$. The effective gap $\Delta_{eff}$ can be used in a 1D traction-displacement law, supplying a straightforward 3D formulation. However, there is no proof that a correct coupling and realistic results are obtained with this approach. In other works, 3D complete models were formulated for various applications like the adhesion of carbon nanotubes (Jiang, 2010), fracture propagation in graded materials (Zhang and Paulino, 2005) or indentation problems (Salehani and Irani, 2018).

In this work, a simplified version of the 3D coupled cohesive laws found in the previous literature is introduced. The adopted traction-displacement relationship is

$$T_i = \Delta_i \frac{\phi_i}{\delta_i^2} \cdot \exp\left(\sum_j -\frac{\Delta_j^2}{\delta_j^2}\right) \tag{2}$$

where $\phi_i$, $\Delta_i$ and $\delta_i$ are the work of separation, the crack gap value and the characteristic length (i.e. the gap value corresponding to the maximum traction), respectively, in the direction $i = [I, II]$, where I indicates the normal direction and II the transverse direction, and $i$ and $j$ are the direction indexes that can assume the values $[x, y, z]$. The energy per unit area $\Delta\gamma_i$ can be defined as $\Delta\gamma_i = \phi_i/A$, where $A$ is the contact area. If the work of separation and the characteristic length are the same for the normal and tangential direction, Eq. (2) becomes

$$T_i = \Delta_i \frac{\phi}{\delta^2} \cdot \exp\left(\frac{-\Delta_{eff}^2}{\delta^2}\right) \tag{3}$$

where Eq. (1) with $\beta = 1$ is used to define $\Delta_{eff}$. The interface stress $\sigma_i$ can now be defined as

$$\sigma_i = \sum T_i / A \tag{4}$$

The simplified cohesive law showed in Eq. (3) is based on several assumptions: the traction and compression behaviour is the same, there is reversibility (which is not the case when damage is present, where the unloading phase is different from the loading one) and there is a softening region. Although Eq. (3) is not suitable to treat mechanical problems where large compressive values occur, the aim of this work is to calculate crack openings where there is little or no compression. This condition should be verified by comparing numerical results with analytical equations.

## 1.1 Theoretical model

To numerically model a continuous body as an elastic membrane, it is important to choose an appropriate discretization criterion. One of the possible approaches is to describe the structure as a grid of points in 3D space connected by 1D bonds forming a network. This approach, first denominated framework method (Hrennikoff, 1941), was introduced in the first half of the past century and has led to the development of numerous discretized models used today (Nukala et al., 2005; Ostoja-Starzewski, 2002; Brely et al., 2015; Costagliola et al., 2018), thanks to its computational advantages. By varying the mechanical properties attributed to the elements, the anisotropic behaviours of heterogeneous materials can be studied. The procedure described in (Valoroso and Champaney, 2006; Zhang and Paulino, 2005) is used to build a grid of *x*-braced elements (Figure 1) to discretize the membrane, which is considered homogenous and linear elastic, with a Poisson's ratio of $\nu = 1/3$, as imposed by plane stress conditions and mesh definition. Other Poisson's ratios can be obtained by introducing 3-node links and other types of meshes (Ostoja-Starzewski, 2002) or by changing the hypotheses made when defining the mechanical properties of the grid.

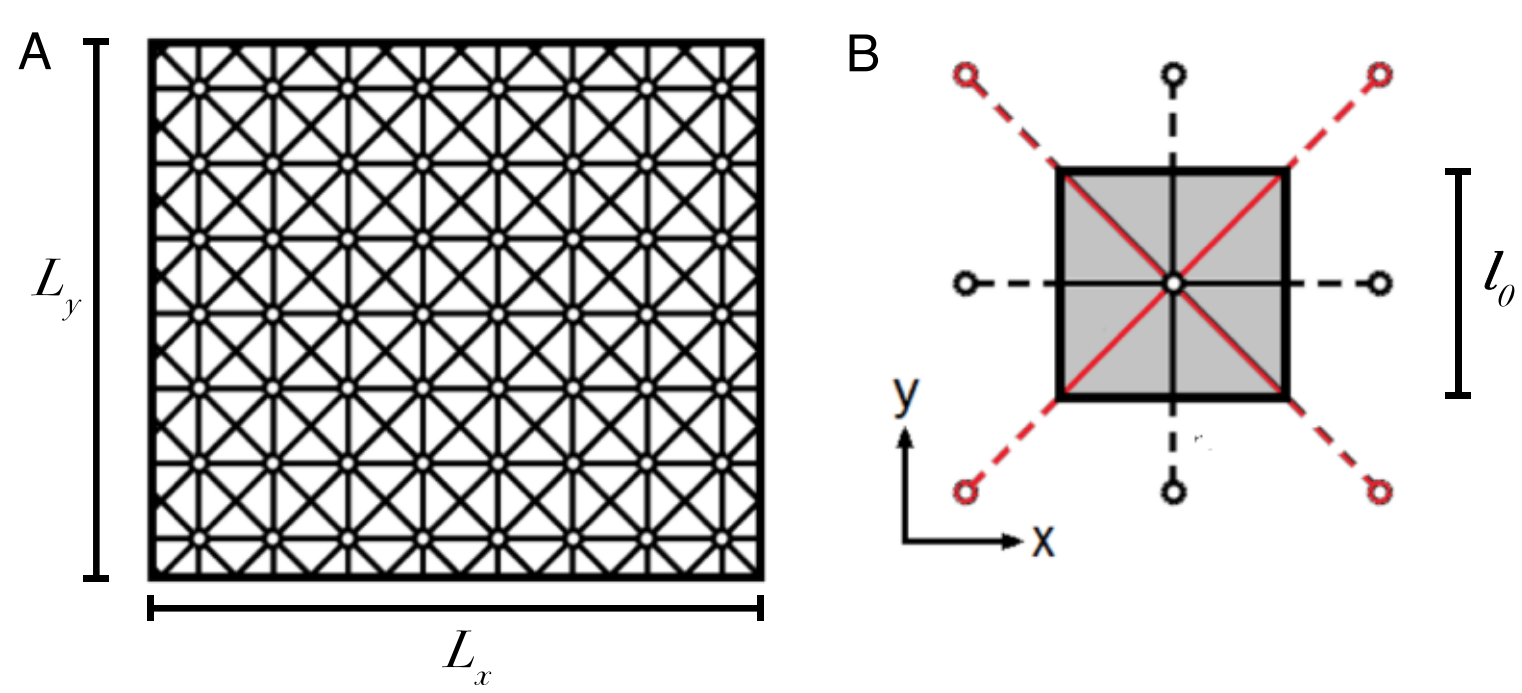

*Figure 1: Membrane discretization: A) Example of a grid made of x-braced squared elements B) Elementary cell. Every node is connected with a truss element to its nearest neighbours (black lines) and next-nearest neighbours (red lines).*

Once the set of points and bonds is defined, a mathematical formulation for the equilibrium equations is needed. In this work, a generalized 3D co-rotational truss formulation is used (Yaw, 2009), i.e. the bonds sustain axial loads only. Given a set of $N$ points $x_i = (x, y, z)$ connected by a grid of $S$ springs, the truss $k$ is defined by its two end points with indexes $p$ and $q$, its cross-section $A$, its initial length $l_0$, and by the constitutive stress-strain equation $\sigma = \sigma(\varepsilon)$ . The internal force vector $\mathbf{Q_i}$ is derived by computing the derivative of the elastic potential energy $U$ with respect to the global displacement vector $v$:

$$\mathbf{Q_i} = \frac{\partial}{\partial v} U \tag{5}$$

The derivative can be rewritten using the chain rule, obtaining:

$$\frac{\partial}{\partial \mathrm{v}} = \frac{\partial}{\partial (l - l_0)} \frac{\partial (l - l_0)}{\partial v} \tag{6}$$

where $l$ is the current length of the truss element. The second term of this differential is given by the direction cosines in the 3D space, which are:

$$n_1 = \frac{x_p - x_q}{l_0} \quad n_2 = \frac{y_p - y_q}{l_0} \quad n_3 = \frac{z_p - z_q}{l_0} \tag{7}$$

The tangent stiffness matrix $\mathbf{K}$, used to linearize the set of equations describing the problem, is defined as

$$\mathbf{K} = \frac{\partial}{\partial v}\mathbf{Q_i} = \frac{\partial^2}{\partial v^2}U \tag{8}$$

Following (Yaw, 2009), the tangent stiffness matrix can be obtained by adding the contributions of the material and the geometric stiffness matrixes ($\mathbf{K_m}$ and $\mathbf{K_g}$). Defining the direction cosine vector as

$$\mathbf{n} = [n_1 \quad n_2 \quad n_3 \quad -n_1 \quad -n_2 \quad -n_3] \tag{9}$$

the two matrices can be written as

$$\mathbf{K_m} = A_k\sigma(\mathrm{\varepsilon_k}) \cdot \ \mathbf{n}^{\mathrm{T}}\mathbf{n} \tag{10}$$

$$\mathbf{K_g} = \frac{A_k}{l_k}\frac{\partial\sigma}{\partial(\varepsilon_k)}\begin{bmatrix} \mathbf{I_3} & -\mathbf{I_3} \\ -\mathbf{I_3} & \mathbf{I_3} \end{bmatrix} \tag{11}$$

where $\mathbf{I_3}$ is the third rank identity matrix. The internal force vector is given by:

$$\mathbf{Q_i} = \sum_k^S A_k\sigma(\mathrm{\varepsilon_k}) \cdot \ \mathbf{n}^{\mathrm{T}} \tag{12}$$

The external force vector $\mathbf{Q_e}$ contains the components of the external load acting on the system. The displacement vector is then updated using the equation

$$\boldsymbol{u} = \boldsymbol{u} + \left(\boldsymbol{K_m} + \boldsymbol{K_g}\right)^{-1}(\boldsymbol{Q_e} - \boldsymbol{Q_i}) \tag{13}$$

The procedure is completed when $\|\mathbf{Q_e} - \mathbf{Q_i}\| < \boldsymbol{\varepsilon}_{\mathbf{conv}}$, where the value of the parameter $\boldsymbol{\varepsilon}_{\mathbf{conv}}$ in the convergence criterion is chosen according to a preliminary convergence test. If the problem is applied to a linear elastic medium, the mechanical constitutive property is

$$\sigma(\varepsilon_k) = E_k \cdot \frac{\Delta l_k}{l_{0_k}} \tag{14}$$

Substituting (14) in Eq. (10), (11) and (12) gives

$$\mathbf{K_m} = K_k \Delta l_k \cdot \ \mathbf{n}^{\mathrm{T}} \mathbf{n} \tag{15}$$

$$\mathbf{K_g} = \frac{K_k}{l_k} \Delta l_k \begin{bmatrix} \mathbf{I_3} & -\mathbf{I_3} \\ -\mathbf{I_3} & \mathbf{I_3} \end{bmatrix} \tag{16}$$

$$\mathbf{Q_i}_k = K_k \Delta l_k \cdot \ \mathbf{n}^{\mathrm{T}} \tag{17}$$

where $E_k$ is the Young’s modulus of the truss member, $K_k = A_k E_k / l_{0_k}$ is its stiffness and $\Delta l_k = \epsilon_k l_{0_k}$ is its elongation.

### 1.2 Numerical procedure

The numerical procedure to solve the system of coupled non-linear equations in matrix form described above is applied using an algorithm based on the Newton-Raphson method. The solution is obtained by linearizing the force vector using a total Lagrangian formulation, as described in (Yaw, 2009; Limkatanyu et al., 2013; Nishino et al., 1984). The algorithm must consider both the contribution of the elastic energy (relative to the deformation of the membrane) and of the adhesive energy (relative to its detachment at the interface). The former is calculated using the co-rotational formulation presented above; the latter is considered by adding to the tangential stiffness matrix ***K*** the Jacobian matrix of the chosen traction-displacement law (Eq. (2)). To simulate the behaviour of the membrane up to total delamination, displacement-control loading conditions are used. The discretization step $\Delta u$ is controlled by an auxiliary algorithm which analyses the convergence speed of the process and varies $\Delta u$ accordingly. The algorithm is implemented in C++. The Armadillo library (Sanderson and Curtin, 2016), OpenBLAS and LAPACK (Dongarra et al., 1993) are used for the linear algebra implementation. The algorithms provided by the superLU library (X. S. Li, 2005) are used to solve Eq. (12). The simulations are run on the OCCAM HPC cluster (Aldinucci et al., 2017) at the Physics department of the University of Torino.

# 2 Validation

Two known cases are considered to validate the numerical procedure, namely single tape peeling and axisymmetric peeling of a membrane.

## 2.1 Tape single peeling

A single peeling test compatible with the hypotheses of Kendall's theory (Kendall, 1975) is considered. The peeling force can be written as

$$F = EL_y t \left[ \cos(\theta) - 1 + \sqrt{\left(1 - cos(\theta)\right)^2 + \frac{2\Delta\gamma}{tE}} \right] \tag{18}$$

where $\Delta\gamma$ is the adhesive energy per unit area, $\theta$ is the pulling angle, $E$ is the Young's modulus of the tape and $L_y$ and $t$ are the width and the thickness of the tape respectively. The ratio $R = \Delta\gamma/\,(tE)$ determines the two, "soft" or "rigid", tape regimes ($R$>>1 or $R$<<1, respectively). Equation (18) is valid adopting the approximation that the stress is concentrated at the peeling line, so that there is no elastic energy stored in the attached section of the tape.

In the case of the numerical model, if $R \ll 1$ regions of the membrane far from the peeling line slip due to the elastic force that exceeds the adhesion force, so that the assumptions of Kendall's theory break down. This effect increases for smaller $\theta$ angles.

As an example, we consider a membrane with the following geometric and mechanical parameters: $L_x = 8\ \mathrm{mm}$, $L_y = 1\ \mathrm{mm}$, $t = 0.01\ \mathrm{mm}$, $E = 1\ \mathrm{MPa}$, $\nu = 0$. The membrane is discretized using square elements of side length $l_0 = 0.1\ \mathrm{mm}$. The adhesive energy is $\Delta\gamma_i = 50\ \mathrm{kPa} \cdot \mathrm{mm}$, which is chosen to work in a relatively soft tape regime ($R = 5$). This is consistent with previous works found in literature, including the original work by Kendall (Kendall, 1975; Heepe et al., 2017; Brely et al., 2018). Figure 2 shows the displacement field of a delaminating tape loaded by a peeling force. Numerical results for the peeling force vs the peeling angle, shown in Figure 2, perfectly match those obtained using Eq. (18), thus validating the numerical code in this particular loading case.

Numerical results are obtained for different values of the characteristic length $\delta_i$. Discrepancies between the theoretical equation and the numerical data are observed in two cases: when $\delta_i \lesssim 3l_0$ the resolution of the cohesive zone is insufficient in the mesh zone where delamination is occurring, and oscillating values of the pull-off force are obtained. Instead, when $\delta_i \gtrsim L_i/4$, the entire membrane slides as soon as a load is applied, so that the maximal force is not reached and border effects prevail. To avoid these discrepancies between simulated and calculated results, $\delta_i$ is chosen in all simulations so that $5l_0 < \delta_i < L_i/5$.

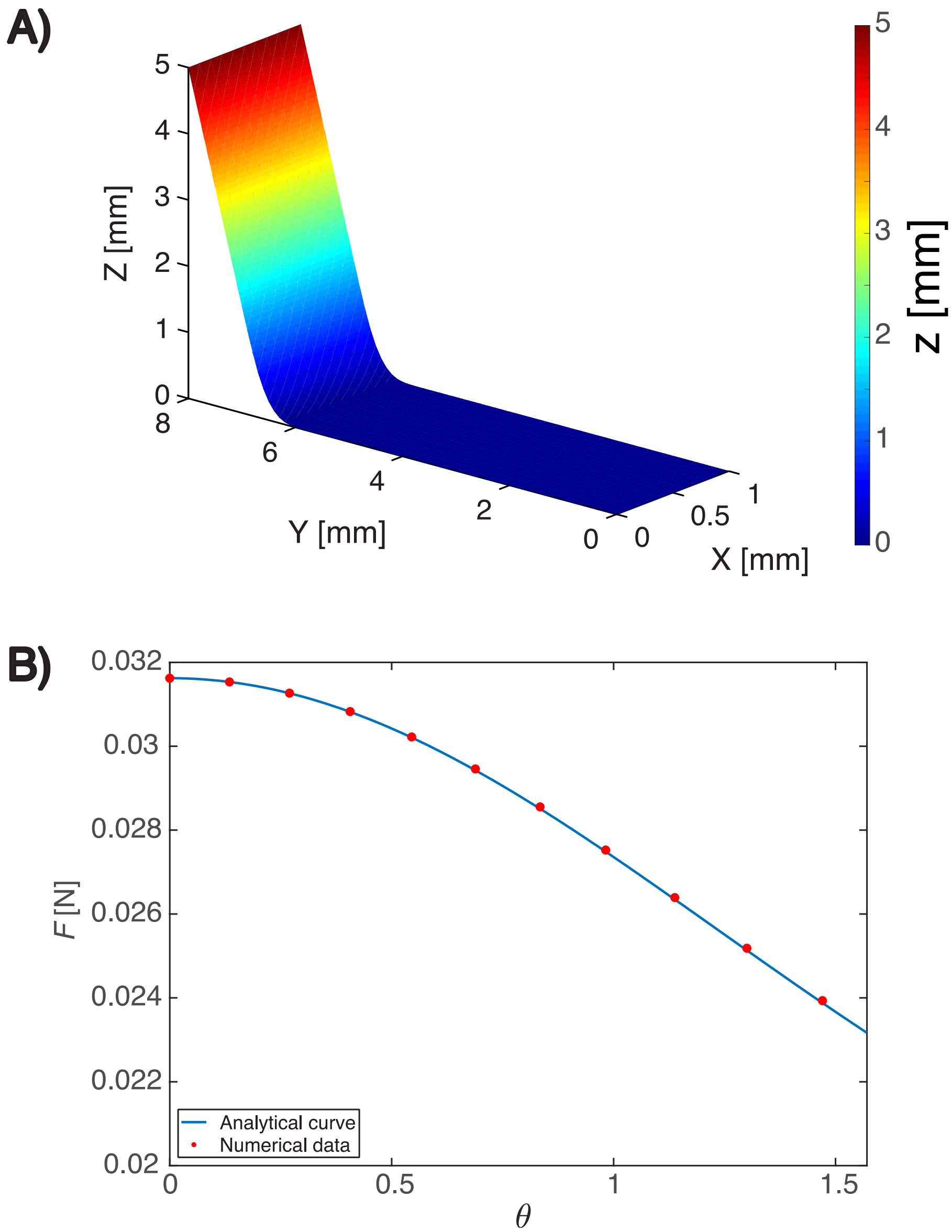

*Figure 2 Numerical simulation of single tape peeling A) 3D displacement map. The colormap indicates the z displacement. B) Numerical vs. analytical prediction of pull-off force vs peeling angle.*

## 2.2 Axisymmetric peeling

Another case which can be treated analytically, and is thus suitable for a comparison with numerical results, is the axisymmetric peeling of a membrane. This problem was solved in (Afferrante et al., 2013) in the case of an infinite membrane attached to a perfectly flat and infinitely rigid substrate. A vertical displacement $u(r = 0)$ is imposed at a single point, and the membrane starts to detach axisymmetrically, as shown in Figure 3. Similarly to the single tape peeling problem described in the previous section, the analytical formulation holds if there is no deformation in the attached section of the tape. As demonstrated in (Afferrante et al., 2013), the force acting on the membrane is

$$F = 2\pi rt \cdot \frac{1}{2} E^* u'(r)^2 \cdot sin\theta \tag{19}$$

where $E^* = E/(1 - \nu^2)$, $\theta$ is the peeling angle, $u(r)$ is the vertical displacement of the membrane as a function of the radius $r$ and $u'(r) = \partial u/\partial r$. Assuming small displacements $\cos d\theta \simeq 1$ and $\sin\theta \simeq u'(r)$, so that Eq. (19) can be rewritten as a differential equation:

$$u'(r)^3 = -\frac{F}{\pi t E^*} \cdot \frac{1}{r} \tag{20}$$

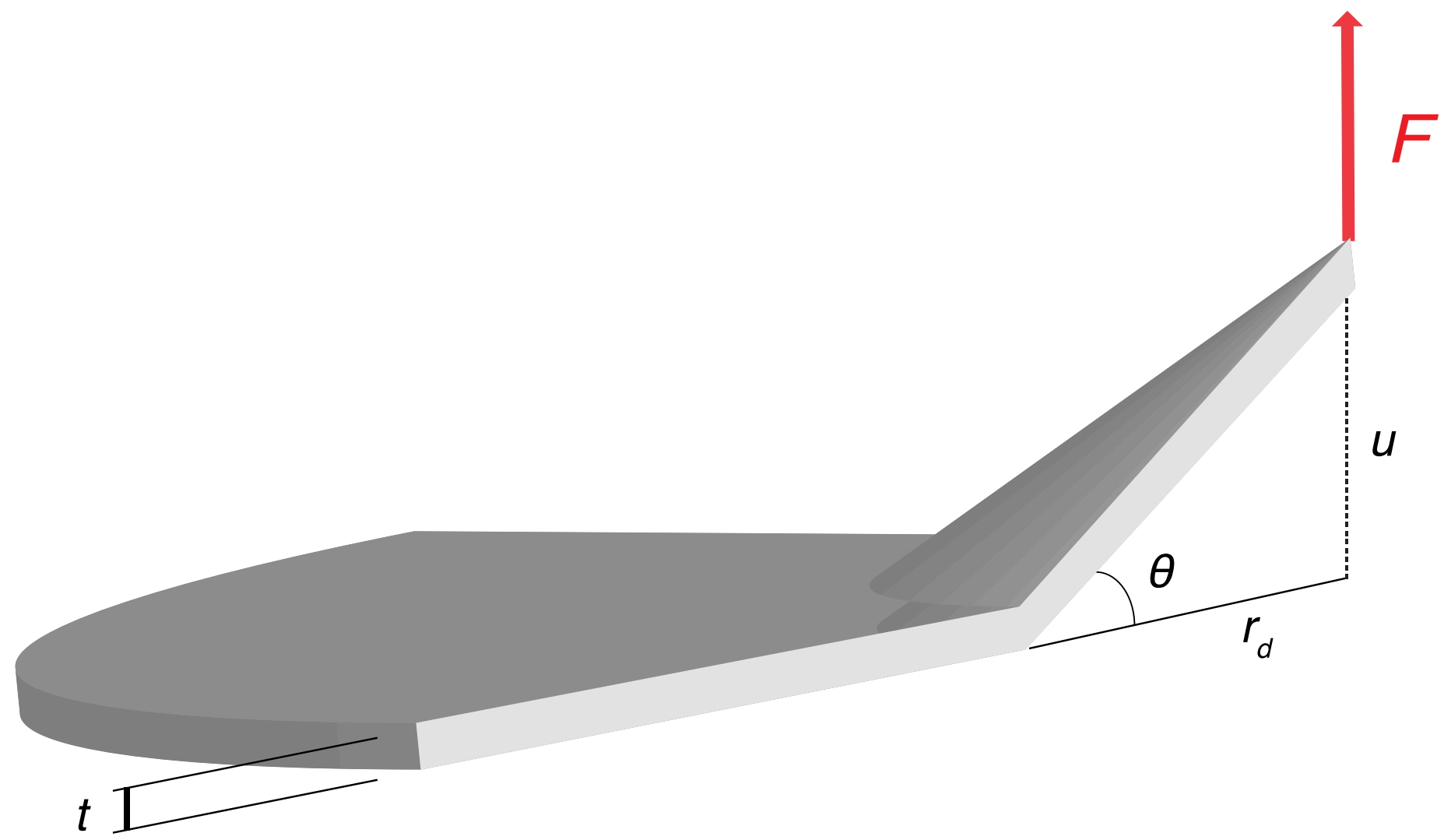

*Figure 3 Graphical representation of the axisymmetric peeling of a membrane (due to symmetry, only one quarter of the membrane is shown).*

To solve this equation, the imposed displacement $u(0) = u$ and the boundary condition $u(r_d) = 0$ are enforced, where $r_d$ is the detached radius. We thus obtain

$$u(r) = \frac{3}{2}\left(\frac{F}{E^* t\pi}\right)^{\frac{1}{3}}\left(r_d^{\frac{2}{3}} - r^{\frac{2}{3}}\right) \tag{21}$$

The load-displacement behaviour of the system is then obtained:

$$F = \frac{8}{27}\pi t E^* \frac{u^3}{r_d^2} \tag{22}$$

This equation can be rewritten to include the adhesive energy of the system. The energy release rate is $G = \frac{1}{2\pi r_d}\left(\frac{\partial U}{\partial r_d}\right)_{\Delta u}$, where $U$ is the total elastic energy. By applying an energy balance criterion, we obtain that $\Delta\gamma = G$, so that Eq. (22) can be rewritten as

$$F \simeq \pi r_d (tE^*)^{\frac{1}{4}} \left(\frac{8}{3}\Delta\gamma\right)^{\frac{3}{4}} \tag{23}$$

where the radial displacement, the circumferential strain and the circumferential stress are assumed to be negligible.

Axisymmetric peeling is modelled numerically as follows. Simulations are performed for a membrane of $L_x = L_y = 1\ \text{mm}$, $E = 0.5\ \text{MPa}$, $t = 1\ \mu\text{m}$. Before load application, the membrane is considered flat and fully adhered to the substrate. Once loading and delamination begin, the detached radius $r_d$ is measured at the point where the maximum delamination load occurs, i.e. it is chosen so that $u(r_d) = \delta$, where $\delta$ is the characteristic length introduced in Eq. (3). Thus, from Eq. (4), the maximum interface stress $\sigma_i$ is $\sigma_{max} = \sigma_i(r = r_d)$.

Figure 4 shows the comparison between numerical and analytical results for the peeling force and displacement $F$ and $u$ as a function of $r_d$ for different values of the ratio $R = \Delta\gamma/(E^*t)$. Good agreement is found, with small discrepancies due to the simplified hypotheses of the analytical model, e.g. a rigid adhesive interface in the limit of small displacements, while in the numerical model, the interface is deformable and displacements can be large. The discrepancy between numerical and analytical results depends on two parameters: $R$, which determines the compliance of the system, and the characteristic length $\delta$. Results in Figure 4 are plotted for $\delta = 0.01\ mm$ and $R < 2 * 10^{-2}$. The peeling force increases approximately linearly with the displacement of the detached radius, as predicted by Eq. (22). This suggests that the force is directly proportional to the length of the peeling line, which is $2\pi r_d$. It can also be observed, as noted in (Afferrante et al., 2013), that the slope of the $u$ vs $r_d$ curve is constant for a given adhesive energy per unit area $\Delta\gamma$. Since $\theta = \mathrm{atan}(u/r_d)$, this means that the peeling angle does not change during delamination, a result which is already found both in single peeling and symmetrical multiple peeling (Pugno and Gorb, 2009; Pugno, 2011; Brely et al., 2014).

To better understand the influence of $R$ and $\delta$, in Figure 6 we compare simulation results to analytical predictions (using Eq. (21)) for the displacement and stress distributions for $[R = 0.01, \delta = 0.01]$ and $[R = 0.1, \delta = 0.1]$. When interface stresses are concentrated along the peeling line, as in Figure 6.A, there is good agreement between analytical and numerical profiles (Figure 6.B). On the other hand, for softer and more deformable structures, the stresses are distributed over a wider zone around the peeling line (Figure 6.C) , which has two effects (Figure 6.D): first, this leads to a wider process zone, which involves the edges of the membrane from the onset of the pull-off phase, introducing edge effects that do not enable to reach the constant $\theta$ steady-state phase; secondly, the deformation occurring in the delaminated part of the membrane displays a larger variation in $\theta(r)$, so that the calculated elongation of the membrane is larger than the simulated one. These two effects are responsible for the discrepancies between the theoretical and numerical results.

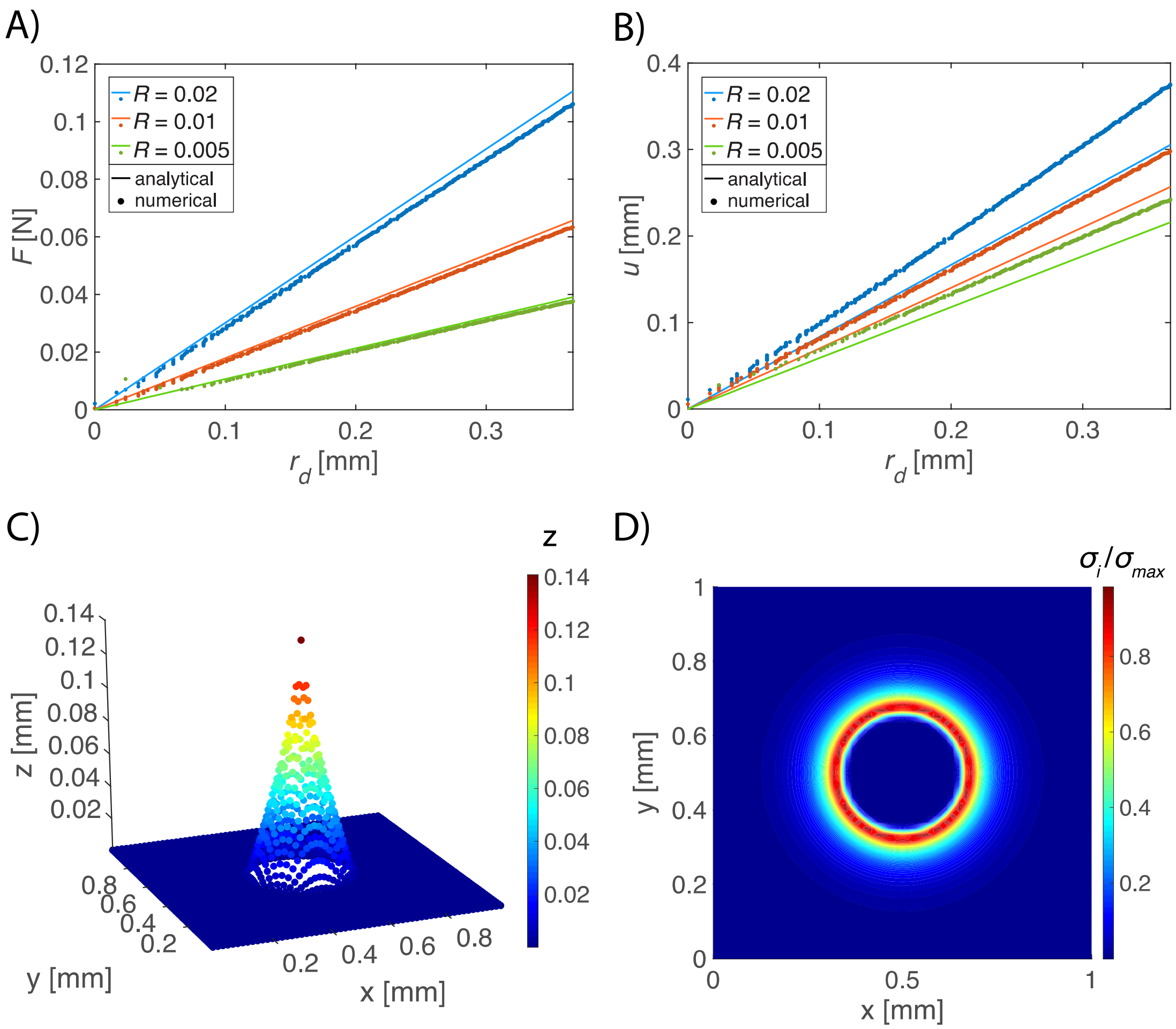

*Figure 4 Axisymmetric peeling of an elastic membrane. A) peeling force vs radius of the detached area for different values of the nondimensional ratio $R = \Delta\gamma/Et$. The continuous lines are the analytical solutions found in (Afferrante et al., 2013), while dots represent the numerical result. B) Imposed displacement vs radius of the detached*

*area for different values of $R$, compared with the analytical solutions. C) Membrane displacement map. The colour map indicates the z displacement. D) Normalized interface stress $\sigma_i$ values at the interface.*

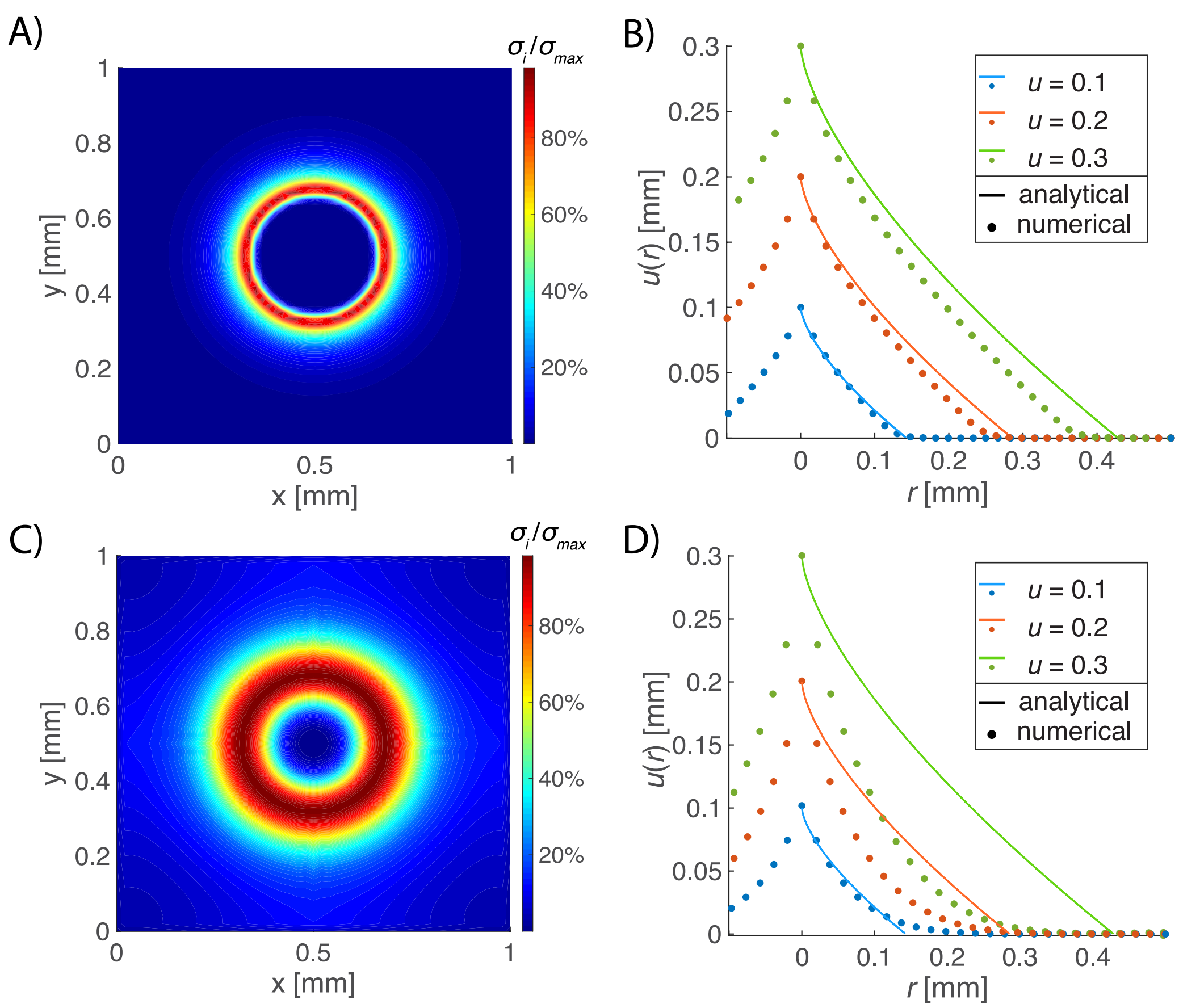

*Figure 5 Force vs detached radius for different values of $R$. Analytical results are compared with numerical simulations for two different values of $\delta$.*

# 3 Results

The peeling behaviour of an elastic membrane depends on numerous parameters: the dimensions of the membrane, its aspect ratio, its Young's modulus, the adhesive energy of the interface, and the loading conditions, i.e. where and how the load is applied. A number of parametric studies are presented in this section to illustrate the model predictive capabilities and to gain insight into the overall behaviour of an adhesive elastic membrane.

## 3.1 Pulling angle and adhesive directionality

We first investigate the effect of the pulling angle $\theta$ on the pull-off force $F$. To do so, we simulate a flat membrane of size $L_x = L_y = 1\ \mathrm{mm}$, thickness $t = 1\ \mu\mathrm{m}$, Young's modulus $E = 0.5\ \mathrm{MPa}$, adhesive energy $\Delta\gamma = 1\ \mathrm{MPa} \cdot \mathrm{mm}$, completely adherent to the substrate. These are typical values for biological adhesive membranes such as, for example, spider disc attachments. The load is applied at a single point located at $y = L_y/2$ (symmetric loading configuration) and at $x = L_x/3$ (asymmetric loading configuration). The configuration is schematically shown in Figure 6.A.

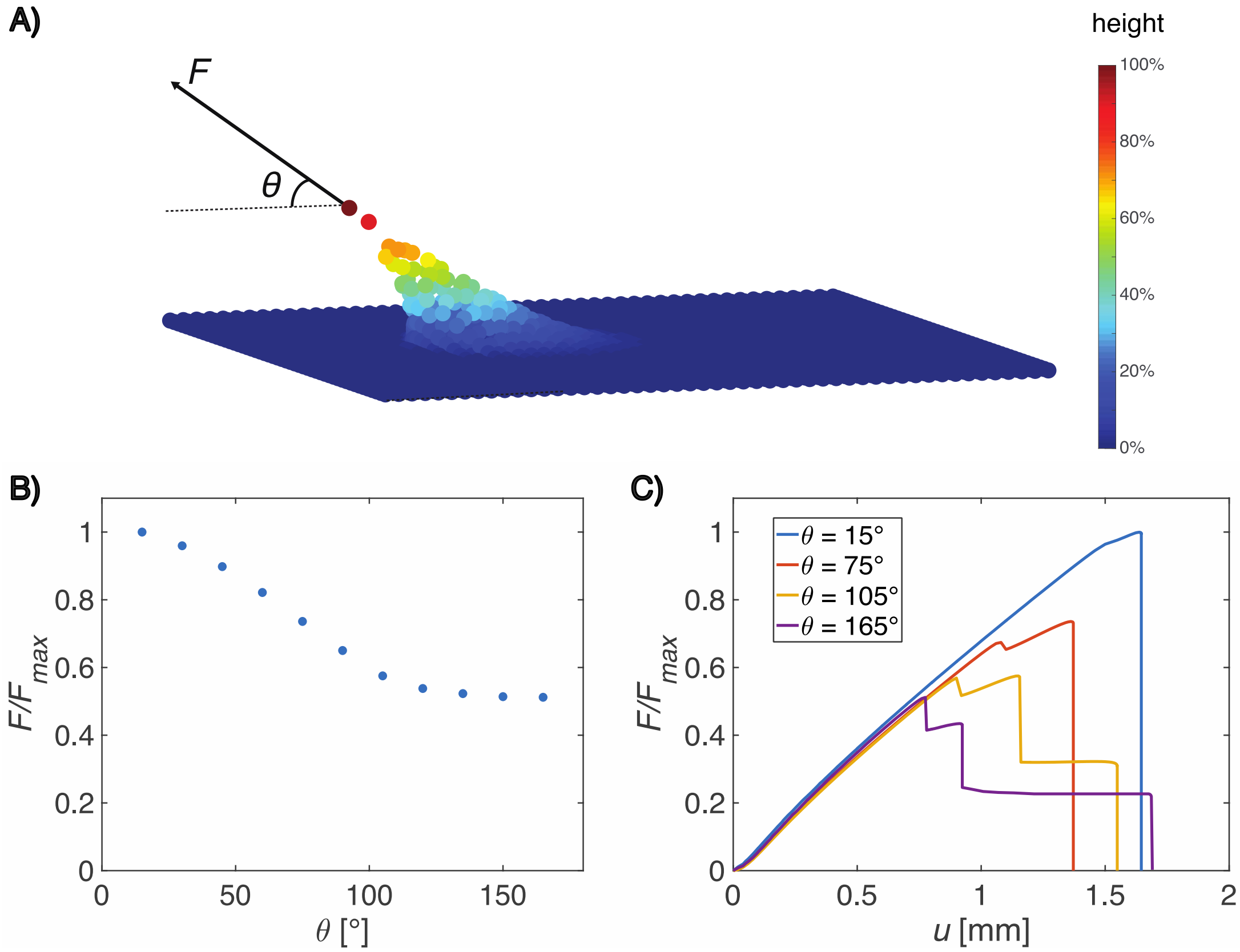

*Figure 6 A) Schematic representation of the simulated case. The membrane is pulled in a single off-centre point by a force F directed at an angle $\theta$. B) Normalized force vs pulling angle $\theta$. C) Normalized force vs displacement for different values of the pulling angle $\theta$.*

Figure 6.B shows the variation of the pull-off force as a function of loading angle. The maximal pull-off force is obtained for $\theta \approx 0°$, while the minimum is obtained in the opposite direction, for $\theta = 180°$. Thus, the membrane displays adhesive directionality and tunability, i.e. there is the possibility of modulating the adhesive force by varying the pulling direction. This is analogous to the Kendall single tape

peeling case. The pull-off force also strongly depends on the location of the pulling point. If the membrane is pulled at its exact centre, results for $\theta = 0°$ and $\theta = 180°$ coincide and the force-angle relationship is symmetrical. Figure 6.B shows the load-displacement relationship at four selected pulling angles for a force applied at $L_x/3$, showing how the pull-off behaviour changes also qualitatively as the angle increases. Each drop in the force coincides with the membrane peeling line reaching the edges of the substrate.

We now focus on the behaviour of the membrane during detachment to better understand how the membrane finite size influences the pull-off force, taking for example data obtained for $\theta = 105°$. Figure 7 shows the interface stress maps of the adhesive interface corresponding to the three force peaks and one of the force drops in the load-displacement plot. The stress distribution corresponding to the peak values indicates that a maximal adhesive force is obtained just before the delamination front reaches a membrane border. After this, a small displacement variation causes a “jump” of the delamination front which is associated to a sudden drop of the pulling force. When continuing to pull the membrane, the curve displays a continuous force increase until another border is reached. After each force drop, the curve increases with a smaller slope than previously.

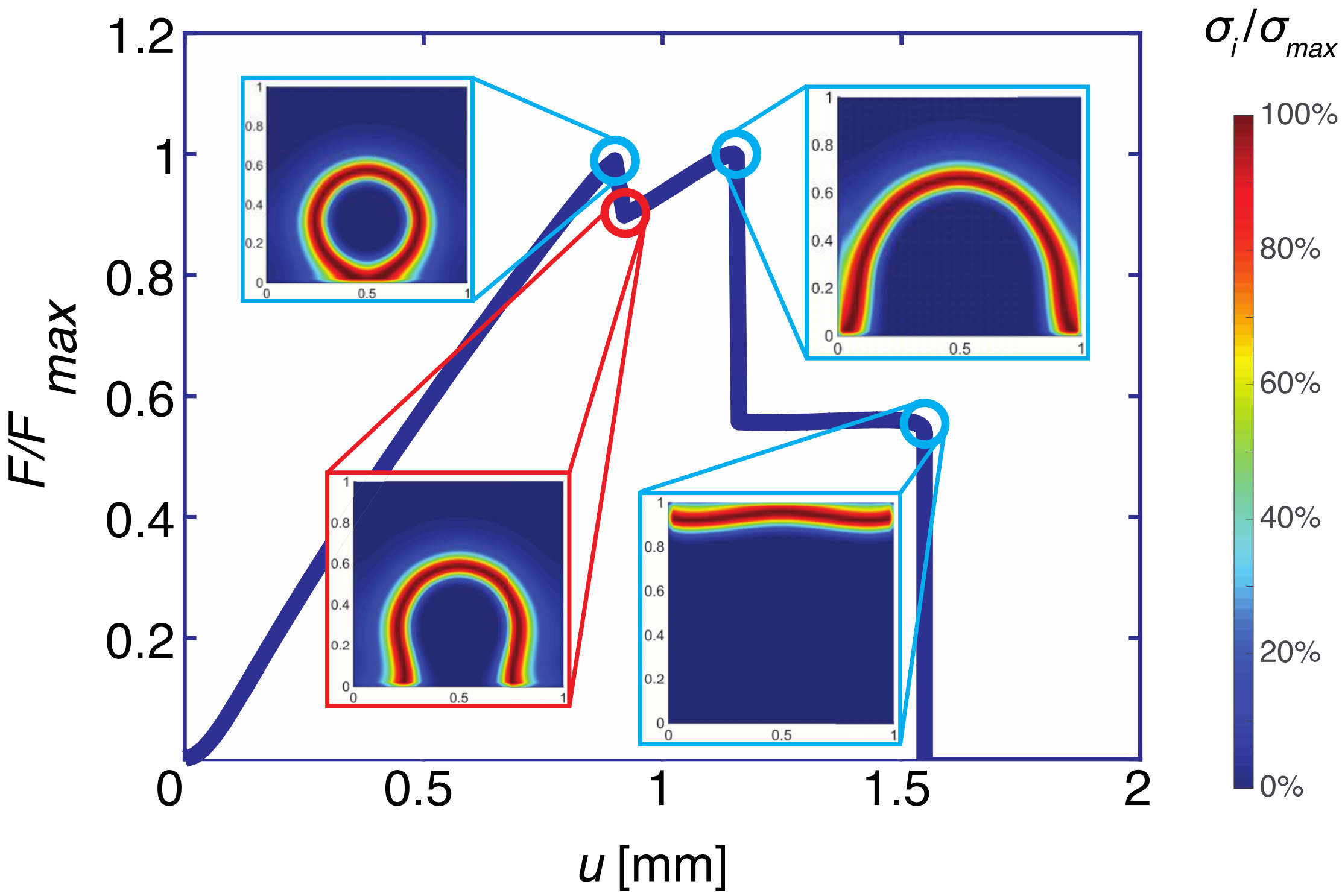

*Figure 7 Normalized force vs displacement for $\theta = 105°$. Normalized interface stress $\sigma_i$ maps highlighting the location of the delamination line in correspondence of some key points of the load-displacement curve. Dark blue indicates points were the interface is not subjected to any stress, which means that the membrane is totally attached or totally detached. Red areas represent points undergoing the maximal stress, i.e. the delamination front ($\Delta u = \delta$, $\sigma_i = \sigma_{max}$).*

A better analysis of the results shown in Figure 6B is now possible. By looking at the different force-displacement curves we see that for $\theta = 15°$, the delamination line reaches all borders almost simultaneously, whilst for $\theta = 165°$ the delamination line reaches the borders at a relatively small load, after which the delamination proceeds with a long tape-like peeling process (at constant load). These behaviours are highlighted by looking at the displacement maps occurring during membrane delamination in the two cases, shown in Figure 8. These plots demonstrate that the numerical model is able to simulate both concave and convex structures in the large displacement regime, which usually gives rise to ill-conditioned numerical problems.

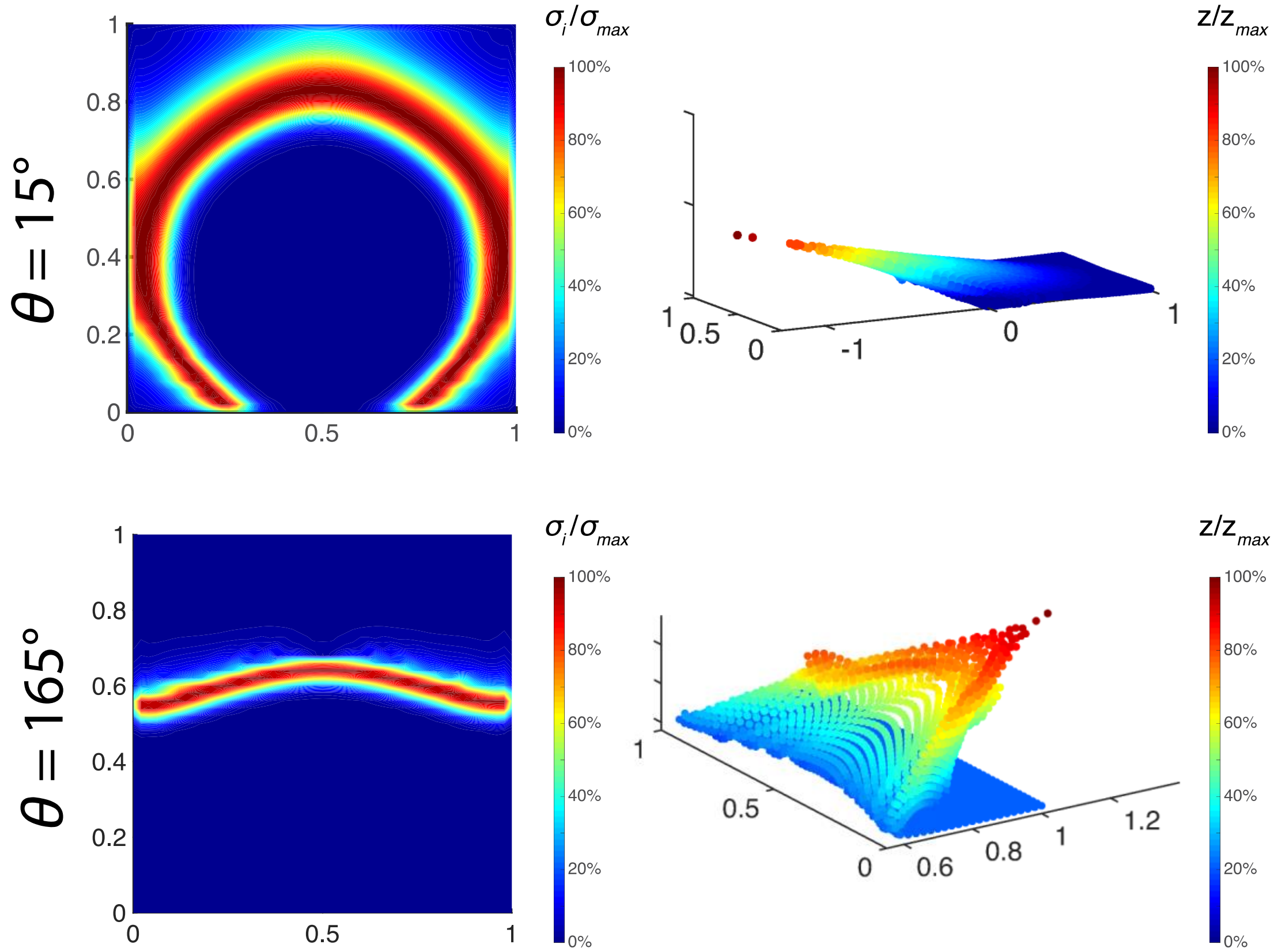

*Figure 8 Interface adhesive stress and membrane 3D displacement plot for the two pulling angles $\theta = 15°$ and $\theta = 165°$. Data for $\theta = 15°$ is taken at the onset of delamination. Data for $\theta = 165°$ shows one of the time steps of the tape-like phase of the delamination. The peeling line, i.e. the length of the delamination front, is much larger in the first case than in the second one. Colours in the interface stress maps on the left show the normalized interface stress, while the colour map in the 3D plots indicates the deformation along the vertical axis of the corresponding portion of the membrane.*

## 3.2 Dependence on the peeling line

To understand how the maximal adhesive force varies with geometrical parameters, it is necessary to determine a correlation between the pull-off force and a global physical quantity. One possibility is to consider the total delaminated area. However, this parameter can be ruled out by looking at Figure 7, where the delaminated area is constantly growing while the force does not vary monotonically. Another possibility is to consider the total peeling line, i.e. the length of the delamination front, which varies non-monotonically during the delamination phase. Results from analysis of the data reported in Figure 6 are shown in Figure 9, where the peeling force is compared to the peeling line length at various points during delamination. The two quantities show a good level of correlation, proving that for a membrane with given mechanical properties, different loading conditions and different geometrical properties affect the shape of the delamination front, whose length in turn determines the pull-off force.

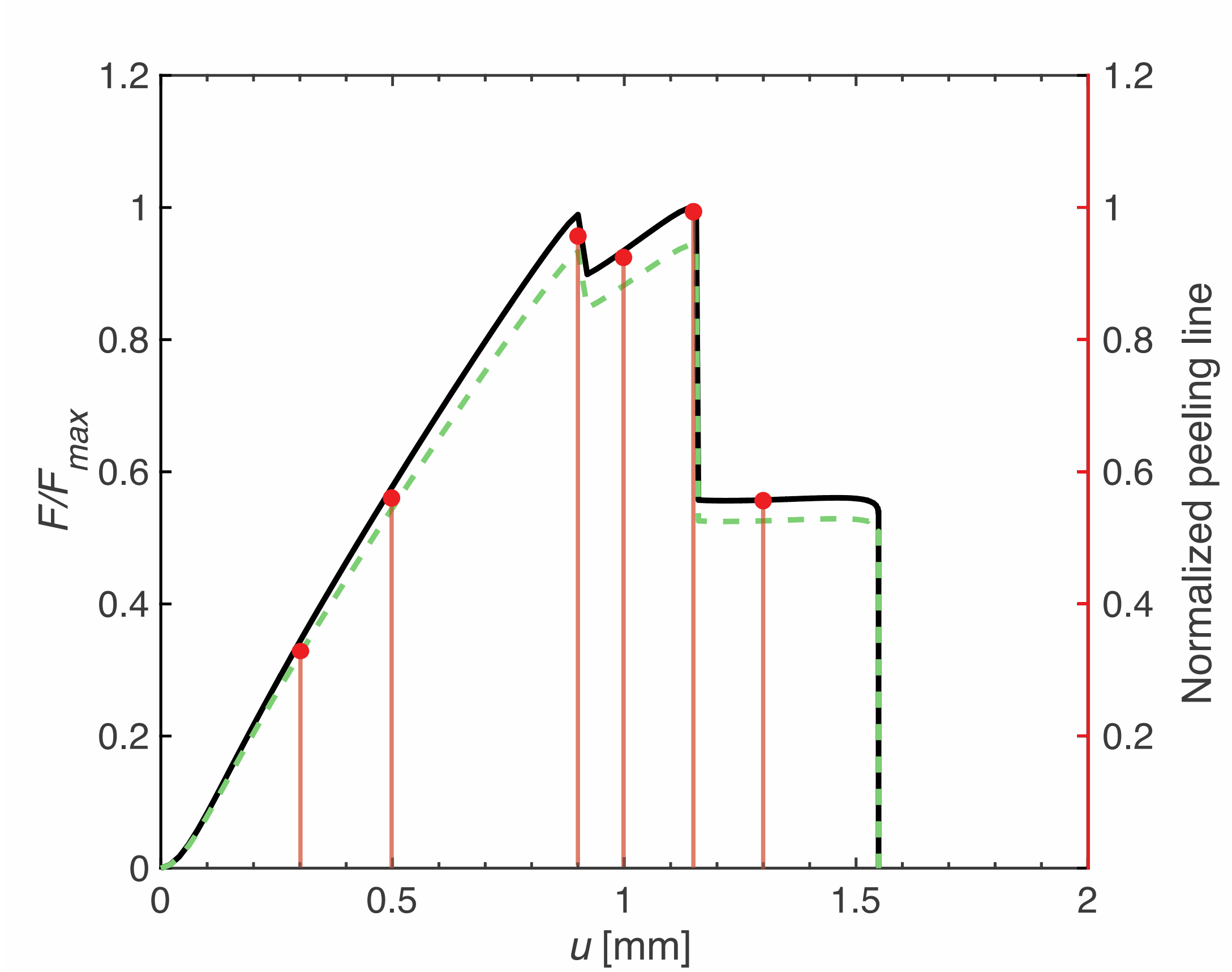

*Figure 9 Normalized force and normalized total peeling line length vs displacement for an elastic membrane pulled at an angle of* $\theta = 105°$. *The two observables display a very good correlation, proving that the length of the delamination front is the main physical quantity which determines the pull-off force during detachment. The dashed line represents the estimated force using Eq. (26).*

To determine the proportionality constant between the pull-off force and the peeling line, we compare the numerically calculated force per unit peeling line $\hat{F}_M$ vs. the peeling angle $\theta$ obtained in the single and double peeling cases. We compare the forces for $\theta \in [0°, 90°]$. $\hat{F}_M$ is symmetrical for $\theta \in [90°, 180°]$. In the double peeling case, the tape is pulled normally to the surface and $\theta$ is the peeling angle instead of the angle of the pulling force. For single peeling, Eq. (18) can be rewritten as (Kendall, 1975):

$$\hat{F}_{\mathrm{SP}} = Et\left[\cos(\theta) - 1 + \sqrt{\left(1 - cos(\theta)\right)^2 + 2R}\right] \tag{24}$$

while for double peeling (Pugno, 2011):

$$\hat{F}_{\mathrm{DP}} = \sin(\theta)Et\left[\cos(\theta) - 1 + \sqrt{\left(1 - cos(\theta)\right)^2 + 2R}\right] \tag{25}$$

Results are shown in Figure 10. It is clear that contrary to the single and double peeling cases, the peeling angle of the membrane is not constant along the whole length of the peeling line. This leads to a variation in the normalized pull-off force, which is found to be intermediate between the single and double peeling cases for small and intermediate angles. Interestingly, for these parameters and peeling angles close to 90°, $\hat{F}$ exceeds the value for the single and double peeling cases, indicating that this configuration realizes a sort of optimum. $\hat{F}_M$ appears to be weakly dependent on $\theta$: we thus compare its values to the analytical prediction for the axisymmetric delamination of a membrane (Eq. (23)), which can provide a theoretical estimation of the peeling line proportionality:

$$\hat{F}_{AX} = \frac{1}{2}\left(\frac{tE}{1-\nu^2}\right)^{\frac{1}{4}}\left(\frac{8}{3}\Delta\gamma\right)^{\frac{3}{4}} \qquad (26)$$

where $\hat{F}_{AX}$ is the force in Eq. (23) divided by the peeling length $2\pi r$. Results are shown in Fig. 10. Clearly, this estimation improves as the pulling angle approaches 90°. In most cases, the discrepancy is small, as can be seen comparing estimated and calculated curves in Fig. 9.

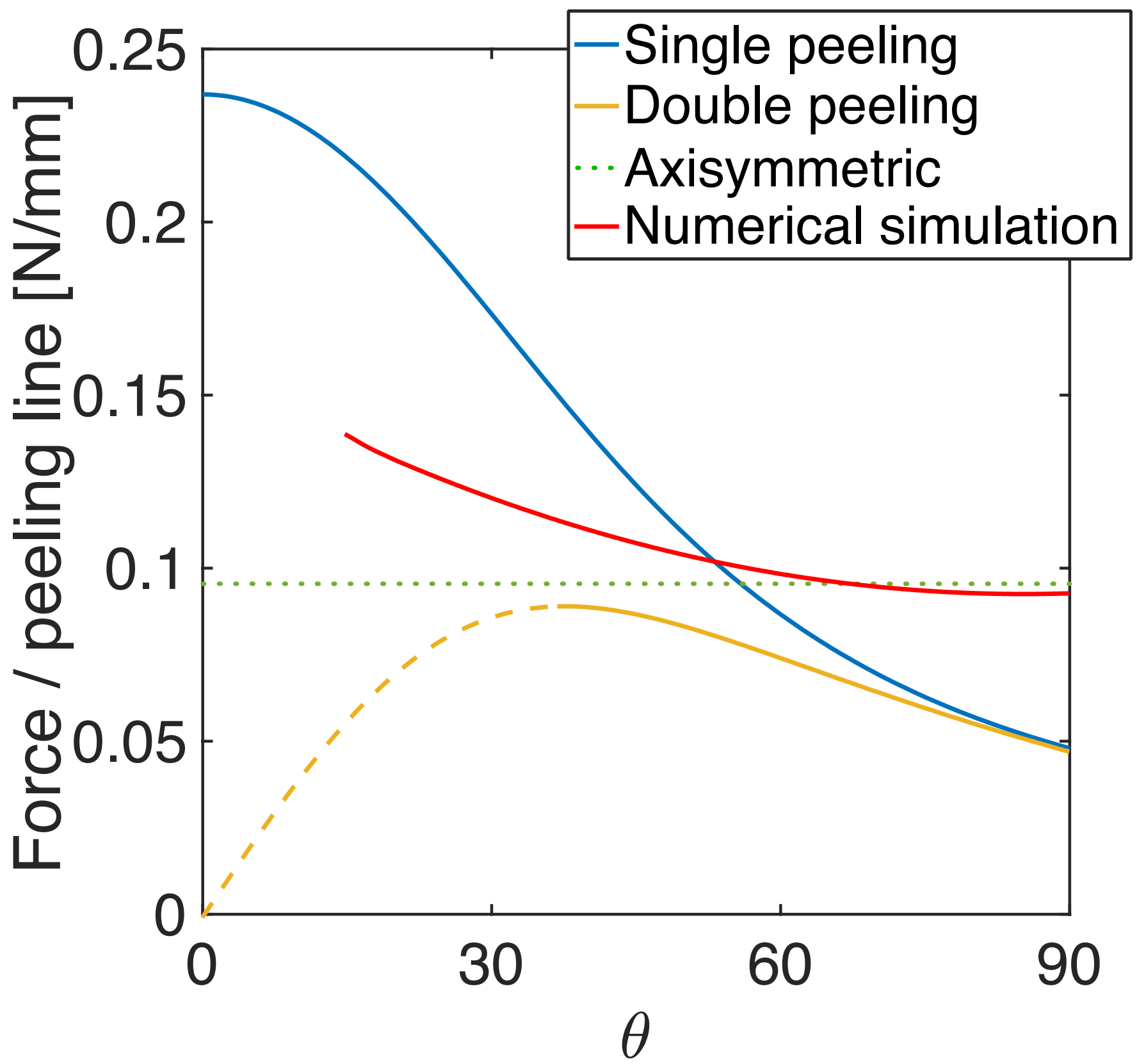

*Figure 10: Force per unit peeling line vs. pulling angle $\theta$ for the three considered cases: membrane delamination, single peeling and double peeling. In the double peeling case, $\theta$ is the angle between the tape and the substrate (the "peeling angle").*

*The dashed part of the curve represents nonphysical values, corresponding to negative initial peeling angles* (Pugno, 2011; Brely et al., 2014). *The dotted line represents estimated force using Eq. (26).*

### 3.3 Dependence on adhesive energy

The adhesive energy, i.e. work of adhesion, is an important mechanical parameter in any adhesion problem. As shown in Eq. (23), the analytical solution of the axisymmetrical peeling force is dependent on the adhesive energy $\Delta\gamma$. In particular, by using the ratio $R$, Eq. (23) can be rewritten:

$$F = \pi r_d t E^* \left(\frac{8}{3} R\right)^{\frac{3}{4}} \tag{27}$$

The dependence of the pull-off force $F$ on $R$ has been discussed in contact splitting problems (Arzt et al., 2003) and in multiple peeling problems (Pugno, 2011; Brely et al., 2014). The behaviour of an adhesive elastic membrane is now studied for different values of the parameter $R$ for $\theta = 90°$. Results are shown in Figure 11. Looking at the force-displacement relationship for different values of $R$, it can be seen that both the strength, i.e. the maximal force, and the extensibility, i.e. the maximal displacement, increase with $R$, but the overall qualitative behaviour is unchanged. The dependence is non-linear and displays a proportionality of $F \propto R^{\frac{3}{4}}$, in accordance with Eq. (27).

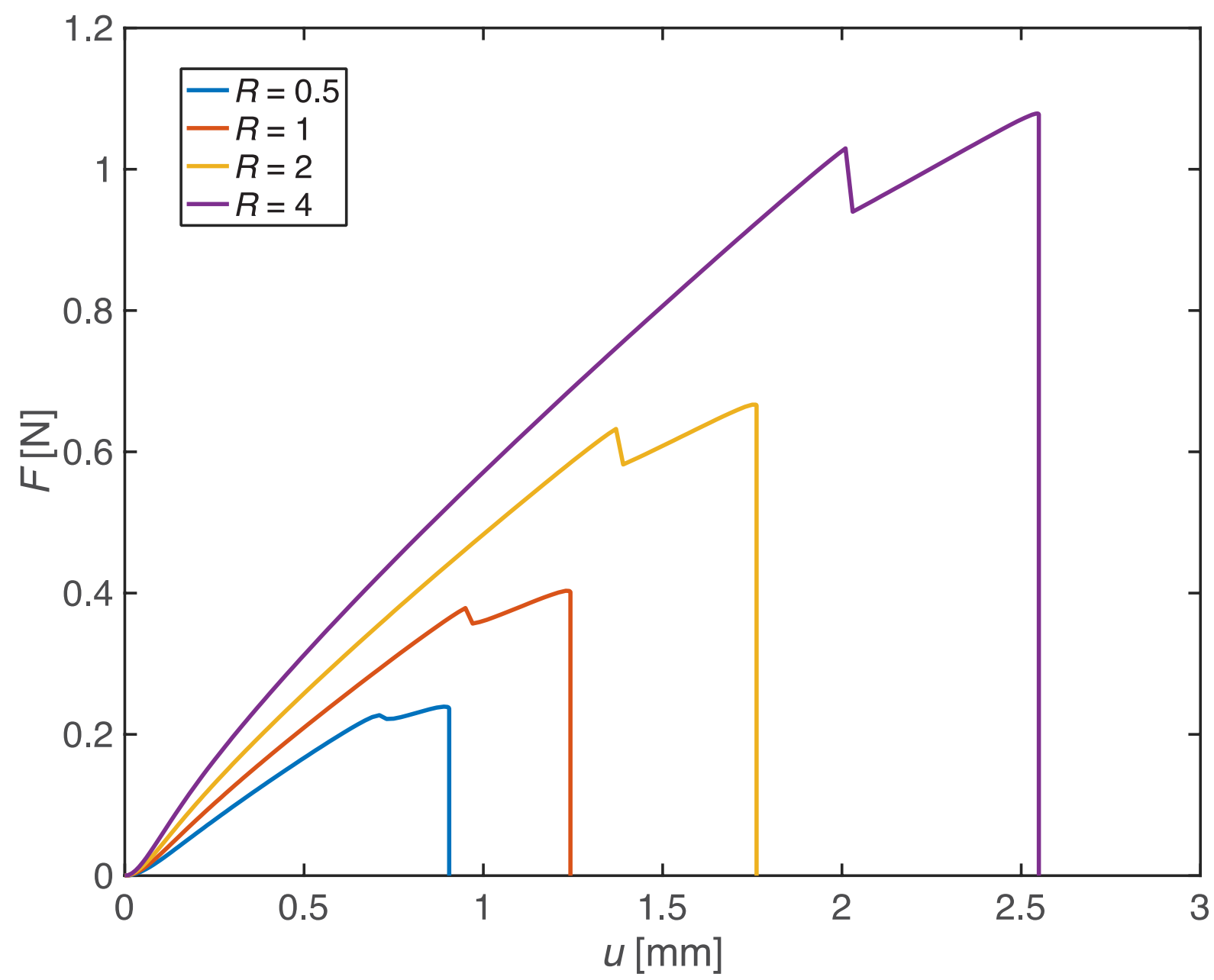

*Figure 11 Force vs displacement for different values of* $R$.

### 3.4 Dependence on membrane aspect ratio

Different loading conditions and mechanical properties have a considerable influence on the adhesive behaviour of the membrane. We now investigate the dependence on the geometrical properties for different pulling angles. To do so, an elastic membrane of area $A = 1\ \mathrm{mm}^2$ is pulled at $\theta = 45°, 90°,\ 135°$ for a force application point located at $L_x/3$ and $L_y/2$. The adhesive energy is $\Delta\gamma = 50\ \mathrm{MPa \cdot mm}$. Simulations are performed for different aspect ratios $\mathrm{L_y}/L_x$. Results are shown in Figure 12.

The pull-off force is strongly dependent on both aspect ratio and loading angle. For a given angle, the pull-off force is maximum for specific aspect ratio values (Figure 12A). For a normal force (90°), two optimal ratios are found when the membrane is slightly larger in width than in length, or vice versa ($L_x \approx 0.75\ L_y, L_y \approx 0.75\ L_x$). If the membrane is too wide or too long, the adhesive force quickly drops down to values $\approx 25\%$ lower than the maximal value for a ratio of $L_y/L_x = 0.5$ and $\approx 35\%$ lower for $\mathrm{L_y/L_x} = 2$. This can again be explained by analysing the force-displacement curves (Figure 12B): when the membrane is too wide or too narrow, the two edges that delaminate first are the front and rear ones, or the two lateral ones, respectively. When this happens, a double peeling phase starts: the force remains relatively constant until total delamination occurs ($L_x = 0.25\ L_y$) or one of the two ends completely detaches and a single peeling phase begins ($L_x\ =\ 4\ L_y$). A similar behaviour is also observed for $\theta = 45°$, but the maximal pull-off force is obtained for an aspect ratio equal to 1. For high pulling angles such as $\theta = 135°$ the membrane starts a single peeling phase at an early stage. In this case, the pull-off force is only dependent from the width of the tape, so larger pull-off forces are reached for wider membranes.

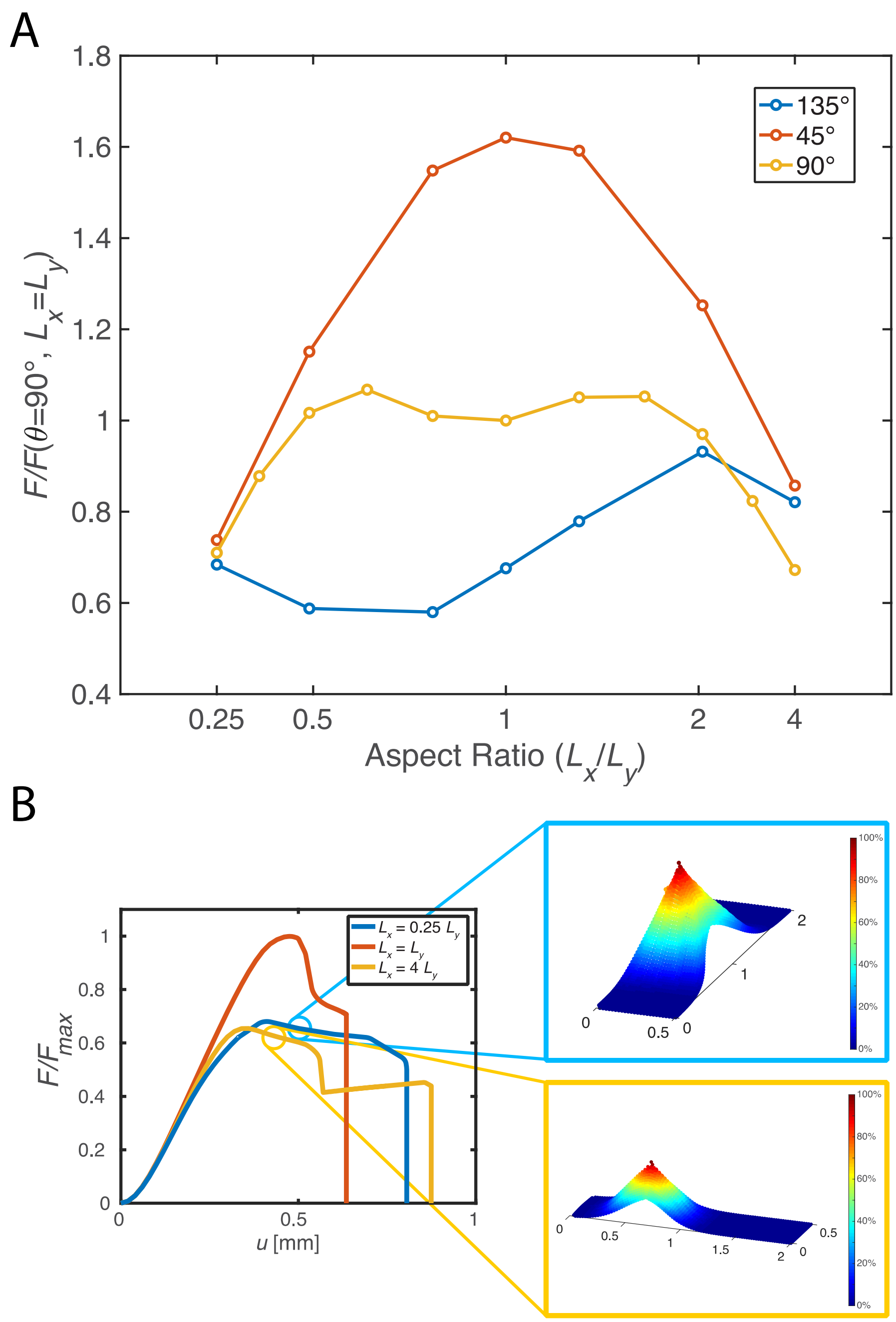

*Figure 12 A) Normalized pull-off force vs the aspect ratio of the membrane for three pulling angles. B) Normalized force vs displacement for three aspect ratios and* $\theta = 90°$. *A 3D displacement plot of the membrane is shown for* $L_x = 0.25\, L_y$ *and* $L_x =$

*4 $L_y$ at the delamination stages corresponding to the indicated points. The colour maps indicate the deformations along the vertical axes. A similar behaviour occurs for corresponding double peeling geometries.*

## 4 Conclusions

In this work, we have presented a new theoretical-numerical model capable of simulating the delamination of elastic membranes attached to a substrate. With the model it is possible to derive total pull-off forces, full 3D displacements and stresses acting on the membrane for oblique pulling forces applied at any point, including in cases when convex regions and ripples develop on the surface. Results have been compared with those obtained by single peeling theory and axisymmetric peeling theory, leading to a validation of the model. The dependence on mechanical and geometrical parameters, such as the aspect ratio of the membrane or the pulling angle, has been highlighted, showing how these are the main factors determining the pull-off force. Moreover, it has been proven that for a membrane of given mechanical characteristics, there is a direct correlation between the pull-off force and the length of the delamination front, i.e. the peeling line. This implies that to maximize the pull-off forces and global adhesion, the membrane should be design in such a way as to maximize how the peeling line (i.e. the maximum stress distributions deriving from membrane deformation) exploit the entire available adhesive area. This can provide inspiration for the design of structured surfaces that allow to exploit this concept for optimized adhesion or anti-adhesion.

This approach can be applied to the study of complex problems with heterogeneous membranes or non-trivial geometries. Further improvements to the model could lead to a better understanding of open mechanical problems in or beyond adhesion. Simulations can be extended to include friction phenomena, using different cohesive laws and interface models, or even fracture phenomena, describing the opening and sliding of a crack interface. Moreover, the versatility of the approach could be exploited to analyse specific biological or bio-inspired problems, such as mussel attachment systems, mushroom-like punches in bioinspired adhesives or octopus suction cups. It is foreseen that more complex membrane constitutive laws, including plasticity or stiffening behaviour, can be easily implemented, thus enabling the reliable simulation of advanced adhesive problems, where the interplay between geometry, structure, material heterogeneity and mechanical constitutive behaviour can lead to unexpected and at times extreme properties.

**Acknowledgements**

FB is supported by the FET Proactive “Neurofibres” grant No. 732344, the COST Action CA15216 “European Network of Bioadhesion Expertise”, by Progetto d'Ateneo/Fondazione San Paolo “Metapp”, n. CSTO160004, and by the Italian

Ministry of Education, University and Research (MIUR) under the “Departments of Excellence” grant L.232/2016. NMP is supported by the European Commission under the Graphene Flagship Core 2 Grant no. 785219 (WP14 “Composites”) and FET Proactive “Neurofibres” Grant no. 732344 as well as by the Italian Ministry of Education, University and Research (MIUR) under the “Departments of Excellence” Grant L.232/2016, the ARS01-01384-PROSCAN Grant and the PRIN-20177TTP3S Grant. Computational resources were provided the Centro di Competenza sul Calcolo Scientifico (C3S) of the University of Torino (c3s.unito.it)